\documentclass[11pt]{article}
\usepackage{graphicx}

\newcommand{\BABARPubYear}    {00}

\newcommand{\BABARConfNumber} {10}
\newcommand{\SLACPubNumber} {8532}

\usepackage{relsize}
\def\babar{\mbox{\slshape B\kern-0.1em{\smaller A}\kern-0.1em
    B\kern-0.1em{\smaller A\kern-0.2em R}}}

\def\ellm       {\ensuremath{\ell^-}}

\def\Kbar  {\kern 0.2em\overline{\kern -0.2em K}{}}

\def\Kzb   {\ensuremath{\Kbar^0}}
\def\KzKzb {\ensuremath{K^0 \kern -0.16em \Kzb}}

\def\Dbar  {\kern 0.2em\overline{\kern -0.2em D}{}}

\def\Dzb   {\ensuremath{\Dbar^0}}
\def\DzDzb {\ensuremath{D^0 {\kern -0.16em \Dzb}}}
\def\Dstar   {\ensuremath{D^*}}

\def\Bz    {\ensuremath{B^0}}
\def\B     {\ensuremath{B}}
\def\Bbar  {\kern 0.18em\overline{\kern -0.18em B}{}}

\def\Bzb   {\ensuremath{\Bbar^0}}

\def\Bs    {\ensuremath{B_s}}
\def\Bsb   {\ensuremath{\Bbar_s}}
 
\def\BzBzb {\ensuremath{B^0 {\kern -0.16em \Bzb}}}

\def\jpsi  {\ensuremath{{J\mskip -3mu/\mskip -2mu\psi\mskip 2mu}}} 
\mathchardef\Upsilon="7107
\def\Y#1S{\ensuremath{\Upsilon{(#1S)}}}

\def\FourS {\Y4S}

\mathchardef\Deltares="7101
\mathchardef\Xi="7104
\mathchardef\Lambda="7103
\mathchardef\Sigma="7106
\mathchardef\Omega="710A
\def\Deltabar   {\kern 0.25em\overline{\kern -0.25em \Deltares}{}}
\def\Lbar {\kern 0.2em\overline{\kern -0.2em\Lambda\kern 0.05em}\kern-0.05em{}}
\def\Sigbar{\kern 0.2em\overline{\kern -0.2em \Sigma}{}}
\def\Xibar{\kern 0.2em\overline{\kern -0.2em \Xi}{}}
\def\Obar{\kern 0.2em\overline{\kern -0.2em \Omega}{}}
\def\Nbar{\kern 0.2em\overline{\kern -0.2em N}{}}
\def\Xbar{\kern 0.2em\overline{\kern -0.2em X}{}}

\def\upsbb {\ensuremath{\Upsilon{\rm( 4S)}\to B\Bbar}}

\def\ev   {\ensuremath{\rm \,e\kern -0.08em V}}
\def\kev  {\ensuremath{\rm \,ke\kern -0.08em V}} 
\def\mev  {\ensuremath{\rm \,Me\kern -0.08em V}} 
\def\gev  {\ensuremath{\rm \,Ge\kern -0.08em V}} 
\def\gevc {\ensuremath{{\rm \,Ge\kern -0.08em V\!/}c}} 
\def\tev  {\ensuremath{\rm \,Te\kern -0.08em V}}
\def\mevc {\ensuremath{{\rm \,Me\kern -0.08em V\!/}c}} 
\def\gevcc{\ensuremath{{\rm \,Ge\kern -0.08em V\!/}c^2}} 
\def\mevcc{\ensuremath{{\rm \,Me\kern -0.08em V\!/}c^2}}

\def\cm   {\ensuremath{\rm \,cm}}

\def\mum  {\ensuremath{\,\mu\rm m}} 

\def\invfb   {\ensuremath{\mbox{\,fb}^{-1}}}
\def\mus  {\ensuremath{\rm \,\mus}}

\def\mus        {\ensuremath{\,\mu{\rm s}}}    
\def\gsim{{~\raise.15em\hbox{$>$}\kern-.85em
          \lower.35em\hbox{$\sim$}~}}
\def\lsim{{~\raise.15em\hbox{$<$}\kern-.85em
          \lower.35em\hbox{$\sim$}~}}

\def\to                 {\ensuremath{\rightarrow}}

\def\pep2{PEP-II}
\def\BF{$B$ Factory}
\def\abf {asymmetric \BF}

\newcommand{\eqref}[1]{Eq.~(\ref{eq:#1})}

\newcommand{\epjc}      [1]  {{Eur.\ Phys.\ Jour.\ C~{\bf #1}}}

\newcommand{\pl}        [1]  {{Phys.\ Lett.\ {\bf #1}}}      

\newcommand{\prl}       [1]  {{Phys.\ Rev.\ Lett.\ {\bf #1}}} 

\def\jetset74   {\mbox{\tt Jetset \hspace{-0.5em}7.\hspace{-0.2em}4}}

\def\dm {\ensuremath{\Delta m_{B^0}}}

\def\delz {\ensuremath{\Delta{ z}}}

\def\dt {\ensuremath{\Delta t}}
\def\fpm {\ensuremath{f_{+-}}}
\def\fzz {\ensuremath{f_{00}}}
\def\ratio {\ensuremath{(b_+^2 \fpm)/ (b_0^2 \fzz)}} 
\def\resof {\ensuremath{f_{reso}(\dt)}}

\def\re_eb {\ensuremath{Re(\varepsilon_B)}}
\def\gevsq  {\ensuremath{\mbox{\,Ge\kern -0.08em V}^2}} 
\def\bch    {\ensuremath{B^{\pm}}}

\long\def\inst#1{\par\nobreak\kern 4pt\nobreak
    {\it #1}\par\vskip 10pt plus 3pt minus 3pt}

\begin{document}
{\pagestyle{empty}

\begin{flushright}
\babar-CONF-\BABARPubYear/\BABARConfNumber \\
SLAC-PUB-\SLACPubNumber
\end{flushright}

\par\vskip 3cm

\begin{center}
\Large \bf Measurement of the time dependence of \boldmath $\Bz\Bzb$
oscillations using inclusive dilepton events
\end{center}
\bigskip

\begin{center}
\large The \babar\ Collaboration\\
\mbox{ }\\
July 25, 2000
\end{center}
\bigskip \bigskip

\begin{center}
\large \bf Abstract
\end{center}
A preliminary study of  time dependence of \BzBzb\ oscillations using 
dilepton events is presented. The flavor of the $B$ meson is determined 
by the charge sign of the lepton. To separate signal leptons from cascade 
 and fake  leptons  we have used a method which combines several  
discriminating variables in a neural network. The time evolution of the oscillations
is studied by reconstructing the time difference  between the decays of the $B$
mesons produced by the   $\Upsilon (4{\rm S})$ decay.
With an integrated  luminosity of  7.7 \invfb\ collected 
on resonance by  \babar\ at the PEP-II asymmetric \BF ,  we measure
 the difference in mass of the neutral $B$ eigenstates, $\dm$, 
to be $(0.507\pm 0.015\pm 0.022)\times 10^{12}\,\hbar\, s^{-1}$.

\vfill
\centerline{Submitted to the XXX$^{th}$ International
   Conference on High Energy Physics, Osaka, Japan.}
\newpage
}

\begin{center}
\small

The \babar\ Collaboration
\bigskip

B.~Aubert,
A.~Boucham,
D.~Boutigny,
I.~De Bonis,
J.~Favier,
J.-M.~Gaillard,
F.~Galeazzi,
A.~Jeremie,
Y.~Karyotakis,
J.~P.~Lees,
P.~Robbe,
V.~Tisserand,
K.~Zachariadou
\inst{Lab de Phys.\ des Particules, F-74941 Annecy-le-Vieux, CEDEX, France}
% ...

\end{center}\newpage

\setcounter{footnote}{0}

\section{Introduction}
\label{sec:Introduction}

A precision measurement of the $\BzBzb$ oscillation 
frequency is of great importance since it is sensitive to the 
CKM matrix element $|V_{td}|$ and, in combination with knowledge of the 
\Bs\Bsb\ oscillation frequency, provides a stringent constraint 
on the Unitarity Triangle. 

The mass difference $\dm $  between the two mass eigenstates of the 
$\BzBzb$ system may be measured by comparing the rate for
pairs of neutral $B$ mesons to decay with the same $b$ quark flavor
with the rate to decay with the opposite flavor sign  at  the \FourS\  
in the following time dependent asymmetry:
\begin{equation}
\frac{N(\Bz\Bzb)(\dt) - (N(\Bz\Bz)(\dt) + 
N(\Bzb\Bzb)(\dt) )}{N(\Bz\Bzb)(\dt) + (N(\Bz\Bz)(\dt) + 
N(\Bzb\Bzb)(\dt) )} = \cos (\dm\cdot\dt),
\end{equation}
where $\dt$ is the  difference between the two  $B$ meson decay times in the \FourS\ 
center of mass system.
The simplest way to determine the $b$ quark flavor of the decaying neutral
$B$  is to use leptons as tagging particles. By counting the number of
``like'' events $(l^+,l^+) + (l^-,l^-)$  and ``unlike'' events $(l^+,l^-)$, 
a measurement of $\dm$ may be extracted through   the
asymmetry : 
\begin{equation}
A_{obs}(|\dt|)=\frac{N(l^+,l^-) - (N(l^+,l^+) + N(l^-,l^-))}{
N(l^+,l^-) + (N(l^+,l^+) + N(l^-,l^-))}.
\label{basicAsy}
\end{equation} 
\par
The semileptonic (muon or electron) branching ratio of B mesons
is about 20\%. Therefore, the dilepton events useful for this analysis
represent 4\%  of the $\upsbb$ decays. 
In  statistical terms, the dilepton tagging is more efficient than
the semi-exclusive tagging performed at the ARGUS \cite{argus_mixing} and the CLEO \cite{cleo_mixing} 
experiments. 
Moreover the  new asymmetric $B$ factories, like PEP-II, allow
a time-dependent measurement, which is
radically different from measurements of the time-integrated probability 
$\chi_d$ performed at the previous 
$e^+e^-$ colliders operating at the \FourS, where  
$\chi_d=x_d^2/(2\cdot(1+x_d^2))$ and $x_d=\dm/\Gamma_{\Bz}$. 
Previous measurements of the time-dependence of \Bz\Bzb\ oscillations 
have been done by the LEP, SLD and CDF experiments \cite{bosc}.

The present measurement is performed on  events collected by the 
\babar\ detector at the \pep2 \abf\  
between January and June 2000. The corresponding integrated luminosity is 7.7\invfb\ taken 
on the \FourS\ resonance and 1.2\invfb\ taken 40\mev\ below resonance. The \babar\ detector and its 
performance are described elsewhere \cite{BabarPub0018}. The event selection and particle 
identification criteria are described in section \ref{sec:Selection}. The selection of signal 
events and a study of the fraction of events with the wrong flavor tagging ({\em{mistag}}) are 
detailed in Section \ref{sec:Tagging}. The method to determine the time-separation of the two $B$ 
semileptonic decays is explained in Section \ref{sec:Deltat}. Section \ref{sec:DmMeas} shows 
the details of the fit on data and the result of the $\dm$ measurement. A list of cross-checks of the 
result is in Section \ref{sec:Crosschecks}, while the evaluation of 
systematic uncertainties is reported in Section \ref{sec:Systematics}. 

\section{Selection of dilepton events}
\label{sec:Selection}

In this study of the oscillation frequency $\dm$ , the flavor of the $B$ meson at decay is 
determined by the 
sign of leptons produced in semileptonic $B$ decays. To reduce the mistag rate, an 
attempt is made to suppress cascade leptons (produced in $b \to c \to \ell$ transitions). 

\subsection{Lepton identification}
Electron and muon candidates are required to pass the {\em{very tight}} 
selection criteria fully described in \cite{BabarPub0018}.
 Electrons are selected by specific requirements 
on the ratio of the energy deposited in the Electromagnetic Calorimeter 
(EMC)
and the momentum measured in the 
Drift Chamber (DCH), on the lateral shape of the energy deposition in the calorimeter, and on the specific 
ionization density 
measured in the DCH. Muons are identified by the use of the energy released in the calorimeter, as well as 
the strip multiplicity, track continuity and penetration depth in the Instrumented Flux Return. 
The performance of the {\em{very tight}} selection criteria are estimated on data control samples, 
as a function 
of the particle momentum as well as the polar and azimuthal angles. The electron and 
muon selection efficiencies are about 92\% and 75\%, respectively, with pion misidentification probabilities around 
0.3\% and 3\%, respectively. 
Lepton candidates consistent with the kaon hypothesis as measured in the
Detector of Internally Reflected Cherenkov light (DIRC) are rejected. More than 60\% of the kaon contamination
in the muon sample is rejected with negligible effect on lepton identification efficiency. 

\subsection{Background rejection}
Non $B\bar B$ events are suppressed by
requiring the Fox-Wolfram ratio of second to zeroth order moments to be less than 0.4. 

The residual contamination from radiative Bhabha and two-photon events is 
reduced by requiring the 
event squared invariant mass to be greater than 20 (\gevcc)$^2$, the event 
aplanarity to be greater than 0.01, and the number of
 charged tracks to be greater than 4.

Electrons from gamma conversions are identified (see  \cite{BabarPub0018})
and rejected
with a negligible loss of efficiency for signal events. 
Leptons from \jpsi\ decays are identified by pairing 
them with the other oppositely-charged candidates of the same lepton species, selected with looser criteria. 
We reject the whole event if any combination has an invariant mass within 40\mevcc\ of the \jpsi\ mass. 


\subsection{Track quality requirements}
We finally apply selection criteria on the quality of the tracks, in order to improve the $\delz$ reconstruction, 
where  $\delz$ is the difference between 
the decay points of the two $B$ mesons along the beam direction. 
Any lepton candidate must have a distance of closest approach to the nominal beam position in the 
transverse plane, $d_0$,  less than 1 cm, and a distance of closest approach along the beam direction, $|z_0|$,  less  
than 6\cm, at 
least 20 hits in the DCH, at least 4 $z$-coordinate hits in the Silicon Vertex Tracker, 
a momentum range in the center of mass system between 700\mevc\ and 2.5\gevc, a momentum range in the laboratory system between 
500\mevc\ and 5\gevc, and a polar angle in the range between 0.5 and 2.6 radians. We also require the 
total error on $\delz$, computed on an event-by-event basis 
to be less than 175\mum. 
When estimating the event-by-event error, it should be noticed that, due to non-zero flight length 
of the B mesons in the transverse plane, the two leptons do not actually originate from the same 
point in that plane. The total error is therefore the quadratic sum of the tracking error and 
of this additional uncertainty. As reported in Section \ref{sec:Deltat}, 
the non-negligible effect of the track quality requirements on signal efficiency 
yields only a small degradation of the resulting statistical uncertainty in $\dm$.

\subsection{ Selection of the direct dileptons}
\label{sec:Tagging}


The discrimination between direct and cascade leptons  is based on a neural network
which combines five discriminating variables, all calculated in the \FourS\ 
center of mass system: 
\begin{itemize}
\item the  momenta of the two leptons with highest momenta, $p^*_1$ and $p^*_2$;
\item the total visible energy, $E_{tot}$, and the missing momentum, 
$p_{miss}$, of the event; 
\item the opening angle between the leptons, $\theta_{12}$. 
\end{itemize}
\begin{figure}[hbtp]
\begin{center}
\mbox{\includegraphics[height=14cm]{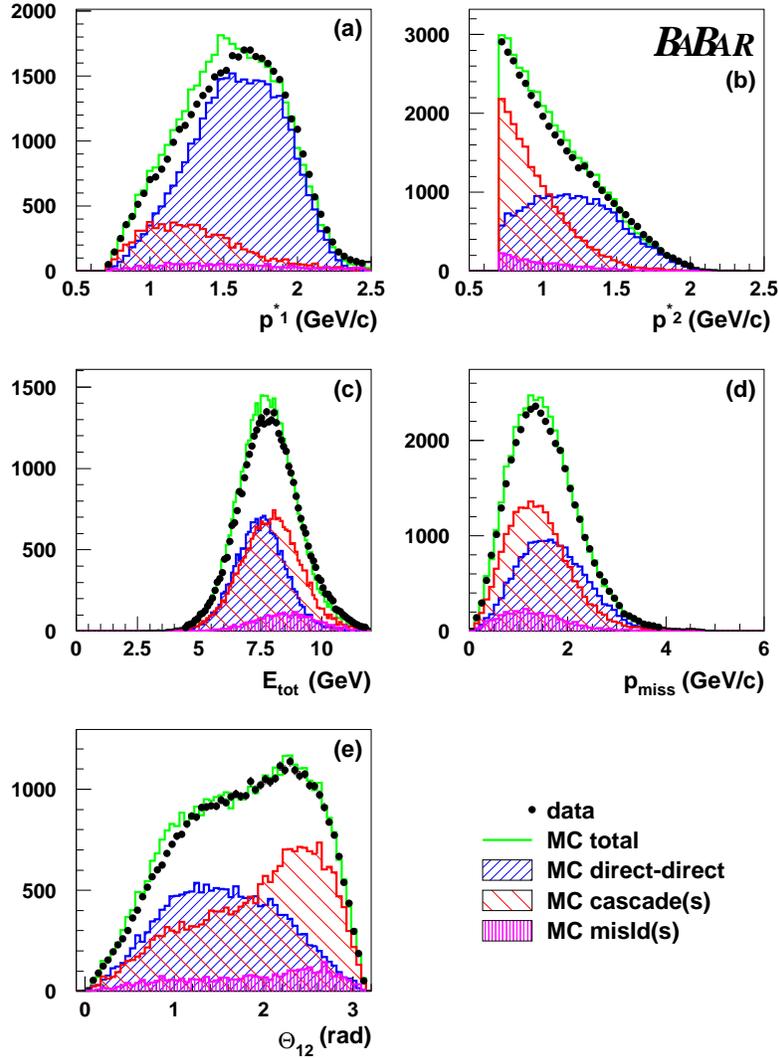}}
\end{center}
\caption{ Distributions of the discriminating variables (a) $p^*_1$, 
(b) $p^*_2$, (c) $E_{tot}$, (d) $p_{miss}$, and (e) $\theta_{12}$,
for data (points) and Monte Carlo (histograms). 
The contributions from direct-direct pairs, direct-cascade pairs, and pairs with one or more fake leptons, are 
shown for the Monte Carlo simulation.}
\label{dis_var}
\end{figure}
The distributions of these variables are shown in Figure  
\ref{dis_var}, for data and Monte Carlo simulation. 
The first two variables, $p^*_1$ and $p^*_2$, are very powerful in discriminating between direct and
cascade leptons. The last variable, $\theta_{12}$, efficiently removes 
direct-cascade lepton pairs coming from the same $B$ and further rejects gamma conversions. 
Some additional discriminating power is also provided by the other two variables. 
The  chosen neural network architecture (5:5:2) is composed of 3 layers, with  2 outputs in the last layer
(one for each lepton). The network is 
trained with 40,000 dileptons from generic \Bz\ and \bch, and the outputs are chosen to 
be 1 and 0 for direct and cascade leptons respectively. 
\begin{figure}[hbt]
\begin{center}
\mbox{\includegraphics[height=8.cm]{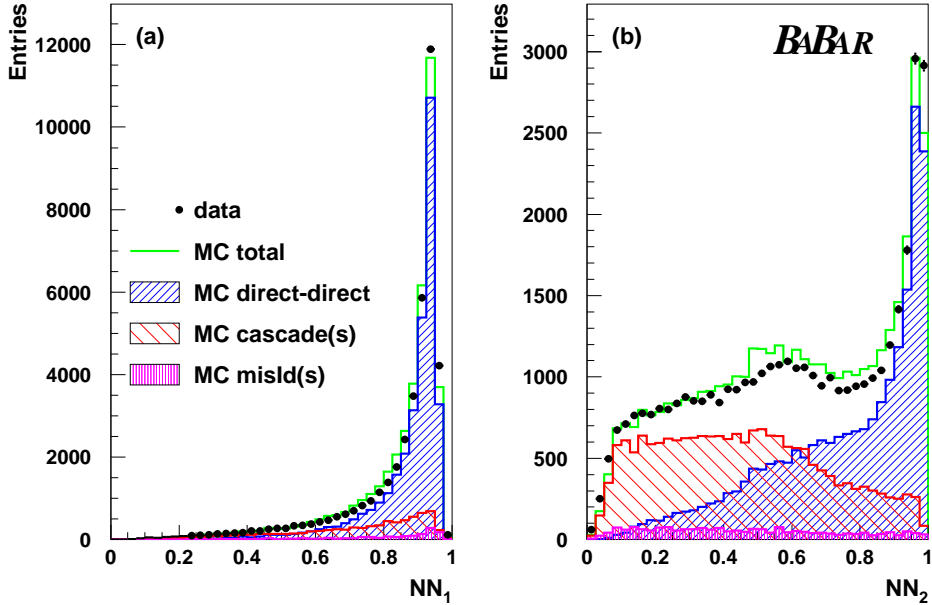}}
\end{center}
\caption{Neural network outputs distributions for the (a) highest and (b) second-highest 
momentum leptons, for data and Monte Carlo simulation. The various Monte Carlo contributions are shown 
separately.}
\label{nnoutput}
\end{figure} 
Figure \ref{nnoutput} shows good agreement between data and simulation 
for the neural network outputs of the two leptons. We require both outputs to be greater than 0.8.

The combined effect of the above cuts gives, from simulated events, signal purity and efficiency of 78\% and 
9\%, respectively. 
The remaining background consists of 12\% direct-cascade events (8\% with the wrong tagging),
5\% $\B\Bbar$ events with one or more fake leptons, 2\% $\B\Bbar$ events with one or more non-prompt leptons, 
a negligible contribution from cascade-cascade events, 
and 3\% from continuum events. The latter was determined in data 
by rescaling the number of off-resonance events that pass the selection 
with the ratio of on- and off-resonance luminosities. 
The total number of selected on-resonance events is 36631 (10742 electron pairs, 7836 muon pairs, and 
18053 electron-muon pairs).


\section{Determination of \boldmath $\Delta t$ }
\label{sec:Deltat}

A determination of the $z$ coordinate of the $B$ decay vertex using only 
the lepton track can be obtained, to first approximation, by taking the $z$ of the 
point of closest approach between the track and the beam spot in the transverse plane.  
This estimator is a fairly good way to determine the $z$ position of the \Bz\ decays 
vertices since the selected direct leptons have rather high momenta. 
However, it is possible to use the two lepton tracks and a beam spot constraint in a simple $\chi^2$ vertex fit 
to obtain a better estimate of the primary vertex of the event in the transverse plane, and to compute 
the points of closest approach of the two tracks to this new point. The corresponding $z$ coordinates 
represent a better approximation of the $z$ coordinate of the $B$ decay vertices, and the corresponding  
$\delz$ resolution function has much reduced tails. For this reason we adopt the latter method to 
compute the $\delz$. 

Further studies show that a requirement on the total error to be less than 175\mum\ reduces the tails of the 
$\delz$ resolution function by a factor four and reduces the signal efficiency by 30\%. 
However, due to the improved resolution, the total statistical uncertainty on \dm\ is degraded only by 3\% 
despite the loss of efficiency.
A two-Gaussian fit to the resulting $\delz$ resolution 
function from simulated dilepton events gives $\sigma_n = 87$\mum\ and $\sigma_w = 195$\mum\ 
for the narrow and wide Gaussian, respectively, and 76\% of the events  
in the narrow Gaussian. 

The time difference between the two $B$ decay times is defined as 
$\dt = \delz/( <\beta\gamma> c)$, with $<\beta\gamma>=0.554$. This approximation neglects the
$B$ meson motion in the $\FourS$ rest frame. In this 
inclusive approach it is not possible to determine the exact boost. Therefore,  
the effect of this shift was studied with Monte Carlo by comparing the 
fitted value of $\dm$ with the true $\dt$ and with $\delz/( <\beta\gamma> c)$. 
This study shows that the effect is negligible compared to the current level of accuracy of 
this analysis.

\section{Fitting procedure}

\subsection{Time dependence of the fraction of mistagged events}
\label{mistag}
\begin{figure}[hbt]
\begin{center}
\mbox{\includegraphics[height=8.cm]{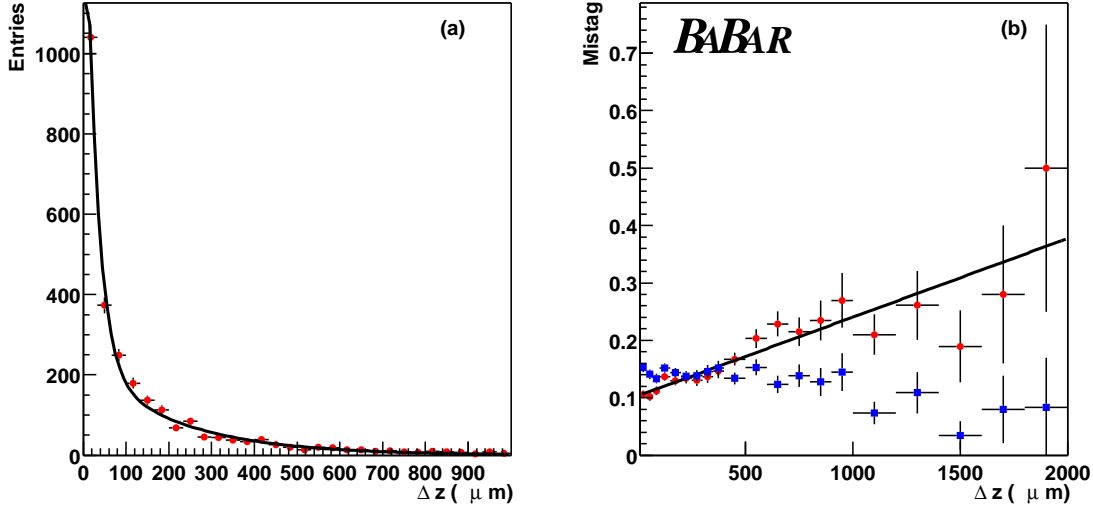}}
\end{center}
\caption{(a) Distribution of \delz\ between direct and cascade leptons that come from the same $B$.
(b) Fraction of mistagged events for cascade leptons coming from the other $B$, as a function of the 
true \delz\ between the two leptons (circles) and the true \delz\ between the $B$ mesons (squares).}
\label{mistagDz}
\end{figure}

Even after a cut on the neural net output, a  non-negligible fraction of events are mistagged (i.e. a true $\Bz\Bzb$ 
pair is tagged as a $\Bz\Bz$ or $\Bzb\Bzb$ pair and vice versa for $\Bz\Bz$ or $\Bzb\Bzb$ events). The 
fraction of mistagged events is directly determined in the fit. 
However we have to take into account that the time dependence of cascade leptons from the same $B$, or from the 
other $B$, are different. In the case of a cascade from a same $B$ 
(Fig.~\ref{mistagDz}(a)), 
we observe a peak at low \delz, due the flight length of 
the charm hadron, which is fitted by an exponential decay.
For the cascade leptons from the other $B$, (circles in  Fig.~\ref{mistagDz}(b)),   
the fraction of mistagged events 
as a function of the true \delz\ between the two leptons shows a linear dependence which
comes from the fact that the  \delz\ measured between the two leptons  contains the additional flight length 
of the charm hadron. The same distribution, determined using the $z$ distance of the true $B$ vertices, is flat to  
first order (squares in  Figure \ref{mistagDz}). Actually, the linear dependence on \delz\ can be explained
by considering the time distribution of the cascade lepton.  Assuming that the flight length of the charmed hadron is 
small compared to that of the $B$, the time distribution of the cascade lepton from the other $B$ can be approximated by:
$$
\eta\, e^{-t/(\tau_B +<\tau_c>)} \simeq \eta\,(1+\frac{<\tau_c>}{\tau_{B}^{2}} 
\cdot t)\,e^{-t/\tau_B},
$$
where $\tau_B$ and $<\tau_c>$ are the $B$ meson lifetime and 
an average lifetime of the $D$ mesons, respectively. 
In the final fit procedure, the linear dependence is taken into account by a free parameter; 
the time shape of  the cascade lepton from a same $B$ is determined from the  Monte Carlo simulation.

\subsection{Measurement of \boldmath $\dm$ }
\label{sec:DmMeas}

The value and statistical error for $\dm$ are extracted with a $\chi^2$ minimization fit to the dilepton asymmetry 
(see Eq. \ref{basicAsy}). The fit function, $A_{fit}(\dt)$,  
takes into account the various time distributions of the dilepton signal ($f^{unmix}(\dt)$, $f^{mix}(\dt)$), 
the cascade lepton and the non-$B\overline{B}$ backgrounds ($f_{other}^{OS}(\dt)$, $ f_{other}^{SS}(\dt)$): 

$$
A_{fit}(\dt) =  \frac { (f^{OS} - f^{SS})\otimes \resof + (f_{other}^{OS}(\dt)-f_{other}^{SS}(\dt))}
 { (f^{OS} + f^{SS})\otimes \resof + (f_{other}^{OS}(\dt)+f_{other}^{SS}(\dt))},
$$

\noindent where $\otimes$ stands for the convolution product with the resolution function $\resof$ (see 
Section \ref{sec:Deltat}), 
and the $f$-functions are expressed in terms of the various signal and background contributions as   

\begin{eqnarray*}
f^{OS}(\dt) & = & f^{unmix}(\dt) \cdot (1 -(1+f_c)\eta_0 ) + f_{mistag}^{SB}(\dt)\cdot f_c \cdot\eta_0 \\
& &+ f^{mix}(\dt) \cdot (\eta_0 + \alpha\dt) \\
f^{SS}(\dt) & = & f^{mix}(\dt) \cdot (1 - (1+f_c)\eta_0 ) +  f^{unmix}(\dt) \cdot (\eta_0 + \alpha\dt), \\
\end{eqnarray*}

\noindent
The signal contributions for unmixed and mixed events are given respectively by  
\begin{eqnarray*}
f^{unmix}(\dt) & = &  \frac{1}{2(1+R)}[ \Gamma^0e^{-\Gamma^0 |\dt|} (1+\cos(\dm \dt)) + 
2R\cdot\Gamma^+ \cdot e^{-\Gamma^+ |\dt|}] , \\ 
f^{mix}(\dt) & = & \frac{1}{2(1+R)}[ \Gamma^0e^{-\Gamma^0 |\dt|} (1-\cos(\dm \dt))], \\
\end{eqnarray*}
where $R$ is proportional to 
the ratio $\ratio$ ($b_+$ and $b_0$ are respectively the semileptonic branching 
ratio of charged and neutral $B$, and $\fpm/\fzz$ is the production ratio of charged and neutral 
$B$ pairs at the $\FourS$). 

The time distribution of direct-cascade events where both leptons originate from the same $B$ is represented 
by $f_{mistag}^{SB}(\dt)= [ <\Gamma^c> e^{-<\Gamma^c> |\dt|}]$, with $<c / \Gamma^c>=60$\mum\ as determined 
from simulated events (Fig.~\ref{mistagDz}). 
The difference in the fraction of direct-cascade events between the cascade 
lepton from the same $B$ and from the other $B$ (due, for 
instance, to the cut on the angle between the 2 leptons) is estimated by the parameter $f_c$, 
determined to be 0.6 from simulated events (Fig.~\ref{mistagDz}). 

The time dependence observed for the mistag fraction of direct-cascade events where the cascade 
lepton comes from the other $B$, as discussed in Section \ref{mistag}, 
is parametrized by a constant term, $\eta_0$, and a slope, $\alpha$. The same functional 
dependences as for signal events, $f^{unmix}(\dt)$ and $f^{mix}(\dt)$, are used. 

The time distributions of the non-$\B\Bbar$ background, $f_{other}^{OS}(\dt)$ and
$ f_{other}^{SS}(\dt)$, and their absolute normalizations, are obtained from off-resonance data. 

Four parameters, $\dm$, $\eta_0$,  $R$, and $\alpha$, are fitted directly to the observed asymmetry. 
The lifetimes of the charged and neutral $B$, $\Gamma^+$ and $\Gamma^0$, are fixed to their 
world average values \cite{pdg98}. 

The off-resonance data is used to measure the fraction 
of the non-$\B\Bbar$ background to be $(0.7\pm0.1)\%$ for same-sign dileptons 
and $(2.2\pm0.3)\%$ for opposite-sign dileptons, respectively. A fit to 
the time distribution of these events yields an effective lifetime equal 
to 130\mum\ and 135\mum\ for same-sign and opposite-sign dileptons, respectively. 

\begin{figure}[hbt]
\begin{center}
\mbox{\includegraphics[height=7.5cm]{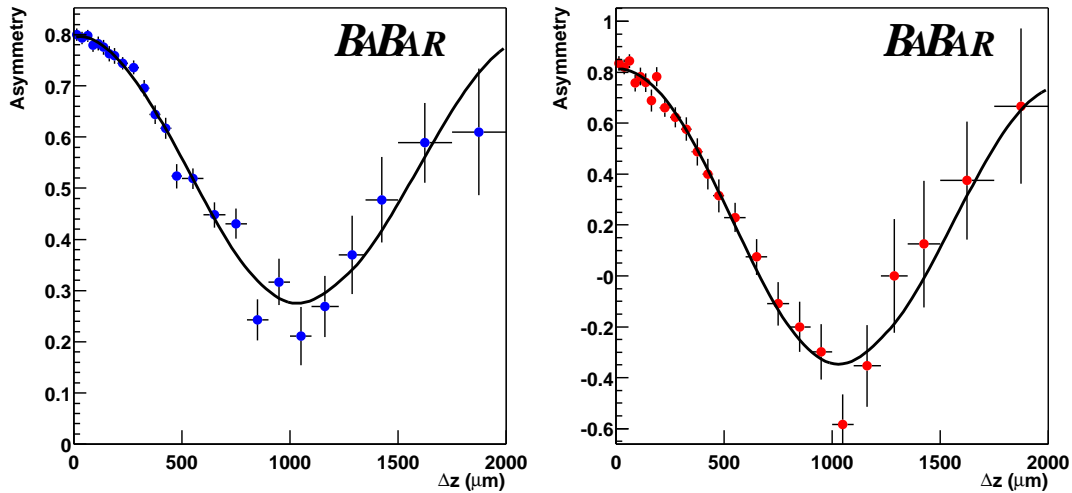}}
\end{center}
\caption{ Distribution of the measured asymmetry $A_{obs}(|\dt|)$ 
between unlike-sign events $(l^+,l^-)$ and 
like-sign events $(l^+,l^+) + (l^-,l^-)$ for (a) the inclusive dilepton sample
and (b) the dilepton sample enriched with soft pions, which is discussed in Section~\ref{sec:Crosschecks}. 
The curve represents the result of the fit.}
\label{asy}
\end{figure}

The fit to the measured asymmetry $A_{fit}(\dt)$ shown in Fig.~\ref{asy} and  
obtained with an integrated  luminosity of  $7.73\, fb^{-1}$, yields the following values: 
$\dm=(0.507\pm 0.015)\times 10^{12}\,\hbar\, s^{-1}$, $\eta_0=0.109\pm 0.004$, $R=1.34\pm 0.11$ and 
$\alpha=(-1.7\pm 3.3)\times 10^{-5}$, with a $\chi^2$ of 20.8 for 21 degrees of freedom. 

\section{Cross-checks and stability of  the \boldmath $\dm$ measurement}
\label{sec:Crosschecks}

\subsection{Enrichment of the neutral {\boldmath $B$} with a soft pion}

In the inclusive approach proposed in this analysis, the final dilepton sample contains both charged and 
neutral mesons $B$ in almost equal proportions. Therefore, the observed oscillation amplitude  
is reduced by the presence of the non-oscillating  
charged $B$. To enrich the $\Bz$ fraction, the direct lepton can be correlated with the soft pion
produced by a  \Dstar$^+$ decay. Charged $B$ mesons can only produce a direct lepton 
and a charged \Dstar\ through the $D^{**}$ decay or through the non-resonant 4 body decay
 $B^- \to D^{*+} \pi^{-} \ellm \overline{\nu}$. The branching fractions of these modes are not perfectly
 measured, but they should represent roughly 10-20\% of the semileptonic decays.

The identification of an event with a soft pion is based on a method proposed by the CLEO 
Collaboration \cite{cleo_mixing}: only tracks with momentum less than 190\mevc\ in the center-of-mass system 
are considered. The 
direction of motion of the \Dstar\ is very close to that of the soft pion (the $D^0$ and the soft pion are
produced almost at rest in the \Dstar\ system) and the energy $E_{\Dstar}^*$ of the \Dstar\ in the 
\FourS\ system is approximated by using the energy of the soft pion $E_{\pi}^*$ in 
the \FourS\  system and the energy of the soft pion $E_{\pi}^{\Dstar}$ in the \Dstar\ system:
 $E_{\Dstar}^* \simeq (E_{\pi}^* / E_{\pi}^{\Dstar})\cdot M_{\Dstar}$. With the four-vector of the lepton 
and the \Dstar, one can compute the missing mass squared $M_m^2$ of the neutrino. 
In the analysis, events are kept if $|M_m^2| \le 1.0 (\gevcc)^2$.

The fit of this sub-sample gives  $\dm=(0.518\pm 0.017)\times 10^{12}\,\hbar\, s^{-1}$, in good agreement with
the value obtained with the dilepton sample. Even though the fraction of events
with the additional  soft pion represents
only 16.5\% of the total dilepton sample, the statistical errors are comparable. While, for the moment, 
this preliminary method constitutes an excellent cross-check, it may later become an alternative approach in its own right.

\subsection{Stability studies}
We have investigated the stability of the fit results against various changes in selection criteria.  
The fit was performed in several ranges of azimuthal angles, as well as for a range of values for 
the cut on the neural network outputs, ($0.6\leftrightarrow 0.9$), 
on the total error of the $\delz$ ($150\mum \leftrightarrow 300\mum$) and for a range in \delz.  
Subsamples composed of only $\mu\mu$, $ee$ and $e\mu$ were also considered. In all cases, 
variations in \dm\ were found to be small or consistent with the nominal value within statistical 
errors. 


\section{Systematic uncertainties}
\label{sec:Systematics}

In this analysis the fraction of mistagged events $\eta_0$ is directly extracted from the
fit of the asymmetry but a time dependence of this component, as well as the fraction of 
misidentified leptons, may induce a bias in the \dm\ determination. These 
effects are corrected by using the time distribution of the mistagged events determined from the Monte Carlo,
and by fitting a slope to the mistag fraction time dependence. The systematic error is determined by  assuming that the 
Monte Carlo corrections are known at the 30\% level. 

A conservative estimate of the uncertainty due to fake leptons is taken to be the difference between the  results of the fit
to Monte Carlo with perfect and simulated particle identification.

\begin{figure}[hbt]
\begin{center}
\mbox{\includegraphics[height=10.cm]{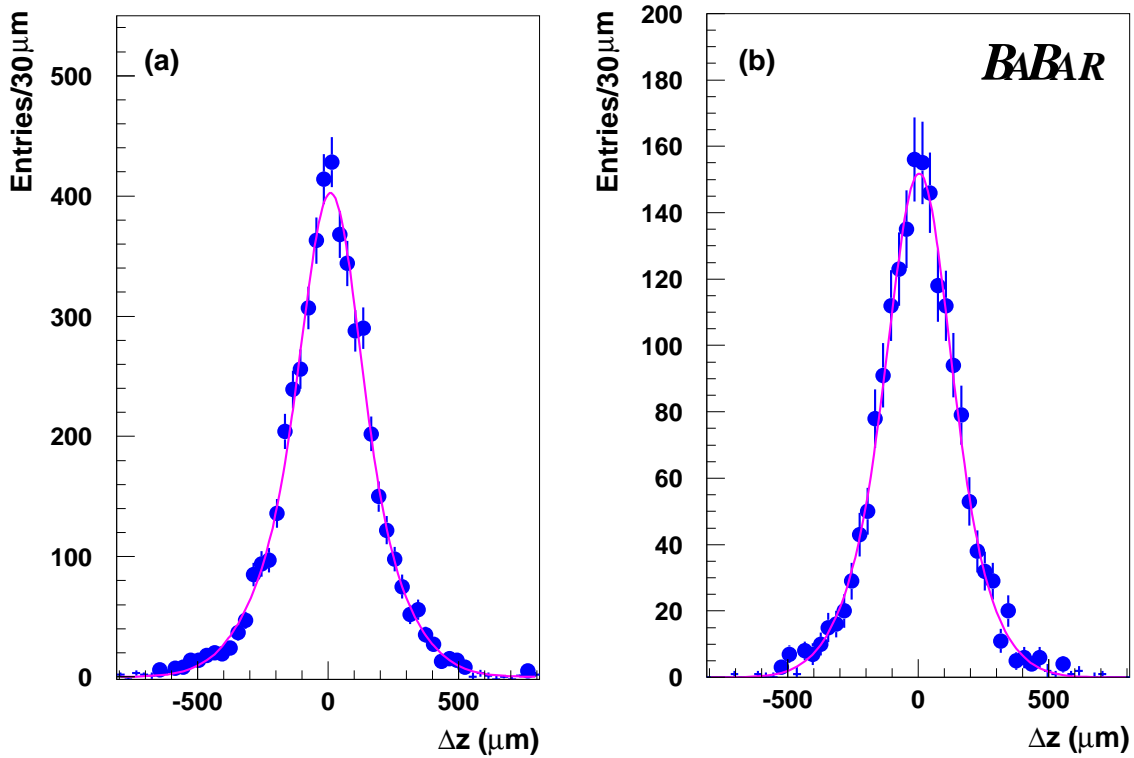}}
\end{center}
\caption{\delz\ distribution  for lepton pairs in the \jpsi\ mass window 
in (a) data and (b) Monte Carlo events. The distributions are fitted with
a sum of two Gaussians. The resolutions for the narrow and wide Gaussians are 101\mum\ and 205\mum\ in data, 
102\mum\ and 184\mum\ in Monte Carlo, respectively.}
\label{jpsi}
\end{figure}

Another important source of systematic errors comes from the determination of the resolution function, which is taken from 
simulated events. To estimate the uncertainty involved in this procedure, we have compared the 
\delz\ resolution between data and Monte Carlo using $\jpsi$ events, where the leptons are known to come from 
the same vertex. 
From this comparison, shown in Figure \ref{jpsi}, 
we estimate an uncertainty on the width of the narrow and wide Gaussians of 
the resolution function of 5\% and 10\%, respectively.

The effect of a charge asymmetry in the identification of the lepton ($\varepsilon^+  \neq \varepsilon^-$) or 
a mistag asymmetry $\eta^+ \neq \eta^-$ on the \dm\ measurement is negligible since the effects 
cancel in the asymmetry. However, the mistag probability $\eta$ may be different for the charged
and neutral $B$.  The impact of such an effect on the \dm\ measurement is negligible 
because the bias is fully absorbed by the parameter $R$, which implies that the 
fitted value of this ratio need not necessarily be unity. 

The list of systematic effects is summarized in Table~\ref{sys_table}. 
The sum of the different contributions gives a total systematic uncertainty of $0.022\times 10^{12}\,\hbar\, s^{-1}$.

\begin{table} [htb]
\begin{center}
\caption{ Summary of the contributions to the systematic uncertainty in \dm.}\vspace{0.3cm}
\begin{tabular}{|l|c|} 
\hline
& $\sigma (\dm )$   \\
Source of systematic uncertainty & $(10^{12}\,\hbar\, s^{-1})$  \\
\hline
\hline
Non-$B\overline{B}$ background  & 0.005 \\ 
\hline
Mis-Identification  & 0.011 \\ 
\hline
Time-dependence of the cascade events  & 0.009 \\ 
\hline
Correction of the boost approximation & 0.001   \\
\hline
y-motion of the beam spot ($\le 20$\mum)  & 0.001 \\ 
\hline
\delz\ resolution function & 0.009  \\
\hline
Tails of the \delz\ resolution function & 0.004  \\
\hline
Time-dependence of the resolution function & 0.006  \\
\hline
Sensitivity to $\Gamma^+$ (PDG 98 $\pm 1\sigma$) & 0.007   \\
\hline
Sensitivity to $\Gamma^0$ (PDG 98 $\pm 1\sigma$) & 0.007  \\
\hline
\hline
Total  & 0.022  \\ 
\hline
\end{tabular}
\label{sys_table} 
\end{center}
\end{table}

\section{Conclusions}
\label{sec:Conclusions}

We  present a preliminary study of the  \BzBzb\ oscillation frequency with an inclusive sample of dilepton events 
corresponding to a total luminosity of $7.73\, fb^{-1}$ collected by the \babar\ experiment.  
We obtain $\dm=(0.507\pm 0.015\pm 0.022)\times 10^{12}\,\hbar\, s^{-1}$. 
The accuracy is already comparable with the current world average. 

\section{Acknowledgments}
\label{sec:Acknowledgments}

We are grateful for the contributions of our \pep2\ colleagues in
achieving the excellent luminosity and machine conditions
that have made this work possible.
We acknowledge support from the
Natural Sciences and Engineering Research Council (Canada),
Institute of High Energy Physics (China),
Commissariat \`a l'Energie Atomique and
Institut National de Physique Nucl\'eaire et de Physique des Particules
(France),
Bundesministerium f\"ur Bildung und Forschung
(Germany),
Istituto Nazionale di Fisica Nucleare (Italy),
The Research Council of Norway,
Ministry of Science and Technology of the Russian Federation,
Particle Physics and Astronomy Research Council (United Kingdom), the
Department of Energy (US),
and the National Science Foundation (US). In addition, individual support 
has been received from the Swiss 
National Foundation, the A. P. Sloan Foundation, the Research Corporation,
and the Alexander von Humboldt Foundation.
The visiting groups wish to thank 
SLAC for the support and kind hospitality
extended to them.


\end{document}